\documentclass{JHEP}
\usepackage{epsfig}
\usepackage{amssymb}
\usepackage[active]{srcltx}

\newcommand{\bea}{\begin{eqnarray}}
\newcommand{\eea}{\end{eqnarray}}

\def\d{{\rm d}}
\def\wt{\widetilde}

\def\wh{\widehat}

\def\th{{\theta}}

\def\a{{\alpha}}

\def\da{{\dot{\alpha}}}

\def\la{\lambda}
\def\U{{\rm U}}

\preprint{ SNUST 041102 \\
{\tt hep-th/0411123} }
\title{Supertwistor Orbifolds:\\
Gauge Theory Amplitudes \& Topological Strings \footnote{ This work
was supported in part by the KOSEF Leading Scientist Grant and the
Alexander von Humboldt Foundation through the Friedlich Wielhelm
Bessel Research Award for SJR, and by the KOSEF Grant
R01-2004-000-10526-0 and the POSTECH BSRI research fund 1RB0410601
for JP.}}
\author{Jaemo Park ${}^{a}$ \,\, \& \,\, Soo-Jong Rey ${}^{b}$\\
~~~~~~~~~~~~~~~~~\\
${}^a$ Physics Department, Pohang University of Science and Technology,
Pohang 790-784 {\rm KOREA} \\
${}^b$ School of Physics, Seoul National University, Seoul 151-747 {\rm KOREA} \\
 \email{sjrey@snu.ac.kr \hskip0.5cm jaemo@postech.ac.kr} }
\abstract{Witten established correspondence between multiparton
amplitudes in four-dimensional maximally supersymmetric gauge
theory and topological string theory on supertwistor space
$\mathbb{CP}^{3|4}$. We extend Witten's correspondence to gauge
theories with lower supersymmetries, product gauge groups, and fermions
and scalars in complex representations. Such gauge theories arise in
high-energy limit of the Standard Model of strong and electroweak interactions. We construct such theories by orbifolding prescription. Much like gauge and string theories, such prescription is applicable equally well to topological string theories on supertwistor space. We work out several examples of orbifolds of $\mathbb{CP}^{3|4}$ that are dual to ${\cal N}=2, 1, 0$ quiver gauge theories. We study gauged sigma model describing topological B-model on the superorbifolds, and explore mirror pairs with particular attention to the parity symmetry. We check the orbifold construction by studying multiparton amplitudes in these theories with particular attention to those involving fermions in bifundamental representations and interactions involving U(1) subgroups.}
\begin{document}
\section{Introduction}
In a remarkable work \cite{witten}, built upon earlier observation
of Nair \cite{nair}, Witten discovered a twistor \cite{twistor}
theoretic reformulation of perturbative super Yang-Mills theory in
terms of topological string theory. Specifically, Witten established
a correspondence between multiparton amplitudes in ${\cal N}=4$
super Yang-Mills theory on ${\mathbb{R}}^{3,1}$ and open string
amplitudes in topological B-model on Calabi-Yau supermanifold
${\mathbb{CP}}^{3|4}$ (Some aspects of Witten's correspondence
relevant for the current work were studied in subsequent works
\cite{related}). Witten's gauge-string correspondence has
successfully reproduced maximal helicity-violating (MHV) amplitudes
at tree-level, and was further extended to a consistent prescription
for reconstructing MHV and non-MHV amplitudes, at tree- \cite{tree}
and one-loop \cite{oneloop} levels, out of MHV sub-amplitudes.

An immediate question is whether, in perturbation theory, it is
possible to extend Witten's gauge-string correspondence to theories with
supersymmetries less than ${\cal N}=4$. This appears not so
obvious since in Witten's formulation the ${\cal N}=4$
supersymmetry was essential; the corresponding supertwistor space
${\mathbb{CP}}^{3|{\cal N}}$ is a Calabi-Yau (super)space only if
${\cal N}=4$ but not for other choices of ${\cal N}$. Another immediate
question is to extend the pure multigluon amplitudes to those
involving fermions or scalars transforming in complex representations. For those transforming in the
adjoint representation, multiparton amplitudes are obtainable
straightforwardly by expanding the ${\cal N}=4$ Yang-Mills theory
amplitudes into component fields. For multiparton amplitudes
involving fermions or scalars in general representations, it is
imperative to consider Yang-Mills theories with lesser or no
supersymmetry. Both questions thus bring us an issue whether a map
analogous to Witten's can be formulated for Yang-Mill theories
with lesser or no supersymmetry, not just at tree level but also
at higher orders in perturbation theory.

In this work, we dwell on this issue and make a step toward the
goal. Our idea is elementary. Take an orbifold action $\Gamma$ that projects
covering ${\cal N}=4$
Yang-Mills theory into a ${\cal N} < 4$ quiver gauge theory. Then, identify
the corresponding operation $\widetilde{\Gamma}$ on supertwistor $\mathbb{CP}^{3|4}$ that projects symmetries and field contents in open string sector of topological B-model to those of the quiver gauge theory. As is well-known, a physical
realization of the ${\cal N}=4$ super Yang-Mills theory is via
worldvolume theory of D3-branes placed on an ambient transverse
$\mathbb{R}^{6}$ in Type IIB string theory. If instead we place
the D3-branes at singular locus of the orbifold
$\mathbb{R}^6/\Gamma$ where $\Gamma$ is an element of discrete
subgroup of SO(6), the worldvolume theory of D3-branes is now
given in terms of gauge theories with lesser or no supersymmetry
\cite{douglasmoore}. Adopting this as a route for
answering the above questions, we will need to understand what the
corresponding topological string theories, if they exist, are and
how they are related to the topological B-model on
$\mathbb{CP}^{3|4}$. We will show below that these topological
string theories are defined on orbifolds of the Calabi-Yau
supermanifold ${\mathbb{CP}}^{3|4}$. More specifically, the
super-orbifolds we propose in this work are the ones in which
discrete subgroup $\Gamma$ of the R-symmetry group SU(4) acts on
fermionic coordinates of the covering supermanifold,
$\mathbb{CP}^{3|4}$. An evident but important point is that such
orbifolding procedure does not violate the super Calabi-Yau
condition that the covering supermanifold obeys. As said, this is a crucial
ingredient for being able to define topological string theory on supermanifold.

Our result suggests that toric super-geometries and super-orbifolds are not just mathematical constructs but bears concrete physical applications in that topological B-models on such superspace are related via Witten's gauge-string correspondence to Yang-Mills theories with appropriate matter contents and supersymmetries. One would consider this relation as opening up a new avenue for supergeometries with interesting physical applications.

It should be straightforward to extend the construction to super-orientifolds --- superspace obtained by orientifolding the Calabi-Yau supermanifold $\mathbb{CP}^{3|4}$. Much like defining super-orbifolds, these super-orientifolds are definable by identifying suitable projection operation as counterpart to orientifolding of the ${\cal N}=4$ super Yang-Mills theory.
We however postpone consideration of super-orientifolds to a separate work
elsewhere.

We organize this work as follows. In section 2, we recapitulate
quiver gauge theories with supersymmetry ${\cal N}=2,1,0$ and work
out the related topological string theories on appropriate
super-orbifolds. We study gauged linear $\sigma$-model description
of such supergeometry and explore mirrors by utilizing
Landau-Ginzburg description. We also comment why it is impossible
to construct ${\cal N}=3$ counterparts. In section 3, to
illustrate utility of such constructions, we study multiparton MHV
amplitudes in quiver gauge theories, with particular attention for
those involving U(1) gauge groups and fermions in complex representations. We construct them from both quiver gauge theories and from topological
B-model on super-orbifolds and confirm agreement between the two results.

\section{Quiver Gauge Theories and Topological String on Super-orbifolds}
Consider a four-dimensional gauge theory with product gauge groups, matter fermions or scalars in complex representations and ${\cal N}=4$
supersymmetries. In string theory context, a class of such gauge theory
arises naturally as worldvolume theory of D3-branes on an orbifold
singularity, which is known as quiver gauge theories
\cite{douglasmoore}. A natural question is whether a topological
string theory corresponding to quiver gauge theory does exist and, if so,
what sort of operation on the supermanifold would be a counterpart
to the orbifold construction. As we shall see, such operation is
indeed identifiable and involves orbifolding fermionic subspace of the
Calabi-Yau supermanifold $\mathbb{CP}^{3|4}$. Thus, under
Witten's gauge-string correspondence, orbifolds of ${\cal N}=4$
super Yang-Mills theory are mapped to topological B-model on super-orbifolds of the $\mathbb{CP}^{3|4}$.

We begin by recalling $N$ coincident D3-branes in ambient
${\mathbb{R}}^{9,1}$. The R-symmetry group on the worldvolume of
the D3-branes is SU(4), which is the spin cover of SO(6), the
rotation group of the space transverse to the D3-brane
worldvolume. The worldvolume theory of the D3-branes is
four-dimensional ${\cal N}=4$ super Yang-Mills theory of gauge
group U($N$) and R-symmetry group SU(4). We shall construct quiver
gauge theories as appropriate quotients of the theory by a
discrete subgroup $\Gamma \in$ SU(4).
\subsection{ ${\cal N}=2$ super-orbifold}

Consider the worldvolume theory on D5-branes at the orbifold
${\mathbb{C}}^2/{\mathbb{Z}}_2$, which is well known to preserve
six-dimensional ${\cal N}=1$ supersymmetry. The theory is of
quiver type and has gauge group $\U(N_1) \times U(N_2)$, 4 scalars
transforming as $({\bf N}_1, \overline{\bf N}_2)$, and 4 scalars
transforming as $(\overline{\bf N}_1, {\bf N}_2)$. Along with
gauge bosons and fermions, they fit to one ${\cal N}=1$ vector
multiplet of $\U(N_1)$, one ${\cal N}=1$ vector multiplet of
$\U(N_2)$, one ${\cal N}=1$ hypermultiplet of $({\bf N}_1,
\overline{\bf N}_2)$, and one ${\cal N}=1$ hypermultiplet of
$(\overline{\bf N}_1, {\bf N}_2)$. Upon dimensional reduction to
four dimensions, we then have ${\cal N} = 2$ vector multiplet of
$\U(N_1)$ and $\U(N_2)$, respectively, plus ${\cal N}=2$
hypermultiplet of $({\bf N}_1, \overline {\bf N}_2)$ and
$(\overline{\bf N}_1, {\bf N}_2)$, respectively. Below, we will
first recapitulate construction of this quiver gauge theory in a
way a direct comparison with fermionic orbifold construction is
transparent. We will then construct the fermionic orbifold
construction explicitly and demonstrate the equivalence.
\subsubsection{Orbifolding Gauge Theory}
Start as the covering theory from ${\cal N}=4$ super Yang-Mills
theory with gauge group U($2N$), and consider the ${\mathbb{Z}}_2$
orbifold action on the theory defined by an element that acts simultaneously in a discrete subgroup of
R-symmetry group SU(4):
\bea \Gamma = \left( \begin{array}{cccc} -1 &&& \\ & -1 && \\ &&
+1 & \\ &&& +1 \end{array} \right) \in \mbox{SU(4)} \label{Gamma}
\eea
and in a discrete subgroup of covering gauge group U($2N$):
\bea \gamma_\Gamma = \left( \begin{array}{cc} +{\mathbb{I}}_{N_1} & \\
& - {\mathbb{I}}_{N_2} \end{array} \right) \in \mbox{U(2N)}.
\label{gamma} \eea
Here, $N_1 + N_2 = 2N$. The large-$N$ superconformal invariance
condition requires Tr$\gamma_\Gamma = 0$, and this is satisfied
only for $N_1 = N_2 = N$.  It is known that planar sector of the resulting quiver gauge theory is idential to the planar sector of the covering theory up to rescaling of gauge coupling parameters \cite{vafaplanar}.

We now identify physical degrees of freedom surviving the orbifold action. Physical modes of the covering theory consist of $2 (2N)^2$ gauge bosons, $6 (2N)^2$ real scalar fields, and $4(2N)^2$ Weyl fermion fields, all transforming in the adjoint representation. Denote also the SU(4) indices as $A = 1, \cdots, 4$. Begin with gauge bosons. Since they are SU(4) singlets, the orbifold conditions are simply
\bea \lambda = + \gamma_\Gamma \lambda \gamma^{-1}_\Gamma,
\label{proj1}\eea
where $\lambda$ represents gauge bosons collectively. This yields
two sets of $N^2$ gauge bosons associated with two disjoint $U(N)$
subgroups of $U(2N)$. The resulting quiver gauge theory has gauge
group $G = \U(N)_1 \times \U(N)_2$. Next, consider gauginos, which
transform as ${\bf 4}$ under SU(4). Those degrees of freedom
surviving the orbifold action of Eqs.(\ref{Gamma}, \ref{gamma})
are
\bea \lambda^A &=& - \gamma_\Gamma \lambda^A \gamma^{-1}_\Gamma
\qquad \mbox{for} \qquad A = 1, 2 \nonumber \\
\lambda^A &=& + \gamma_\Gamma \lambda^A \gamma^{-1}_\Gamma \qquad
\mbox{for} \qquad A = 3, 4 \, , \label{proj2}\eea
where each $\lambda^A$ transforms as a 2-component Weyl spinor.
From Eq.(\ref{proj2}), we obtain 2 adjoint Weyl fermions for
$\U(N)_1$ and 2 adjoint Weyl fermions for $\U(N)_2$, respectively.
We also obtain 2 Weyl fermions transforming as $({\bf N},
\overline{\bf N})$, and 2 Weyl fermions transforming as
$(\overline{\bf N}, {\bf N})$. Finally, consider scalar fields.
Being in antisymmetric representation ${\bf 6}$
\footnote{Recall that $\bf 6$ of SU(4) is also $\bf 6$ of SO(6).}
under the SU(4) R-symmetry group, they are subject to the projection
condition:
\bea \lambda^{[AB]} = \sigma(A) \sigma(B) \, \gamma_\Gamma
\lambda^{[AB]} \gamma_\Gamma^{-1} \, , \label{proj3} \eea
where $\sigma(i)$ equals to $-1$ for $A=1,2$ and to $+1$ for
$A=3,4$. Explicitly,
\bea \lambda^{[12]} &=& + \gamma_\Gamma \lambda^{12}
\gamma^{-1}_\Gamma \qquad \lambda^{[13]} = - \gamma_\Gamma
\lambda^{13} \gamma^{-1}_\Gamma \qquad \lambda^{[14]} = -
\gamma_\Gamma \lambda^{14} \gamma^{-1}_\Gamma
\nonumber \\
\lambda^{[23]} &=& - \gamma_\Gamma \lambda^{[23]}
\gamma^{-1}_\Gamma \qquad \lambda^{[24]} = - \gamma_\Gamma
\lambda^{24} \gamma^{-1}_\Gamma \qquad \lambda^{[34]} = +
\gamma_\Gamma \lambda^{[34]} \gamma^{-1}_\Gamma \,
.\label{scalarcond} \eea
Thus, we have 2 adjoint scalars for $\U(N)_1$, 2 adjoint scalars
for $\U(N)_2$, 4 real scalars transforming as $({\bf N},
\overline{\bf N})$, and 4 real scalars transforming as
$(\overline{\bf N}, {\bf N})$. Putting together, such field
content is precisely that of worldvolume gauge theory for
D3-brane on $\mathbb{C}^2/\mathbb{Z}_2 \times \mathbb{C}$. Notice
that Eq.(\ref{scalarcond}) encodes the orbifold action on
$\mathbb{C}^3$ yielding $\mathbb{C}_2/\mathbb{Z}_2 \times
\mathbb{C}$.
\subsubsection{Orbifolding Topological B-Model}
Next, we would like to understand how the quiver gauge theory can
be mapped to an appropriate topological string theory. To
match with the surviving R-symmetry group in quiver gauge theory,
fermionic subspace of the supertwistor $\mathbb{CP}^{3|4}$ ought
to be projected out accordingly. Since the four-dimensional fermionic subspace have SU(4) symmetry, it is quite evident that
orbifolding the fermionic subspace is a viable operation of doing so. We now show that such operation produces correctly
symmetries and field contents of the quiver gauge theory. We view
this as an evidence that topological B-model on resulting
super-orbifold is the topological string theory we are after.

The parent theory is open string sector of the topological B-model
on $\mathbb{CP}^{3|4}$ \cite{witten}, whose physical states are
described by the $(0,1)$-form ${\cal A}$ on $\mathbb{CP}^3$, which depends on fermionic coordinates $\psi_A$ with $A = 1, 2, 3, 4$ holomorphically. Expanding in powers of $\psi$'s, it is given by \cite{witten}
\bea {\cal A}(Z, \overline{Z}, \psi) &=& \d
\overline{Z}^{\overline{i}} \Big[ A_{\overline{i}} (Z, \overline{Z})
+ \psi_A \chi_{\overline{i}}^A (Z, \overline{Z}) + {1 \over 2!}
\psi_A \psi_B \phi_{\overline{i}}^{[AB]} (Z, \overline{Z})
\nonumber \\
 && + {1 \over 3!} \epsilon^{ABCD} \psi_A \psi_B \psi_C
\widetilde{\chi}_{\overline{i}D} (Z, \overline{Z}) + {1 \over 4!}
\epsilon^{ABCD} \psi_A \psi_B \psi_C \psi_D G_{\overline{i}} (Z,
\overline{Z}) \Big]. \label{A} \eea
The open string sector is described by holomorphic Chern-Simons
gauge theory \cite{wittencs} whose action is given by
\bea I = {1 \over 2} \int_{\mathbb{CP}^{3|4}} \Omega \wedge {\rm
Tr} \Big( {\cal A} \overline{\partial} {\cal A} + {2 \over 3}
{\cal A} \wedge {\cal A} \wedge {\cal A} \Big), \label{CS} \eea
where $\Omega$ is $(3,0 \vert 4)$ form on the Calabi-Yau
supermanifold $\mathbb{CP}^{3|4}$. As explained in \cite{witten}, the
component $(0,1)$-forms $(A_{\bar i}, \chi_{\bar i}^A, \phi_{\bar i}
^{[AB]}, \widetilde{\chi}_{\bar i A}, G_{\bar i})$ match exactly with the
field contents of ${\cal N}=4$ Yang-Mills supermultiplet. It was
further shown that adding D-instanton effects to Eq.(\ref{CS}) on
$\mathbb{CP}^{3|4}$ side yields precisely the perturbative
${\cal N}=4$ super Yang-Mills theory on $\mathbb{R}^4$.

Consider now the holomorphic Chern-Simons theory Eq.(\ref{CS}) but
now with covering gauge group U($2N$). Thus, the Chan-Paton
indices $a,b$ of the component fields appearing in the
$\psi$-expansion Eq.(\ref{A}) runs over $1, \cdots, 2N$. Consider
now phase rotation of the fermionic coordinates, viz.
\bea \psi_A \rightarrow e^{-i \alpha_A} \psi_A, \qquad Z_i,
\overline{Z}_{\bar i} \quad \mbox{intact}. \label{ftrans}\eea
We will choose the phases so that $\alpha_1 + \cdots + \alpha_4 =
0$ modulo $2 \pi$. With such a choice, the measure $\d^3 Z \d^4 \psi$ is
invariant under the rotation Eq.(\ref{ftrans}). It also implies
that the $(3,0|4)$-form $\Omega$ is invariant under the phase
rotation Eq.(\ref{ftrans}) since it is locally given by the
measure $\d^3 Z \d^4 \psi$.

Since $A_{\bar i}$ is inert under Eq.(\ref{ftrans}), in order for
${\cal A}$ to transform as (0,1)-form, the component field
$\chi_{\overline{i}}^A$ ought to transform as
$\chi_{\overline{i}}^A \rightarrow e^{+i \alpha_A}
\chi_{\overline{i}}^A$. Other component $(0,1)$-form fields would
also transform appropriately, which we can read off straightforwardly
from Eq.(\ref{A}). Combined with the aforementioned U($2N$)
Chan-Paton factors, the rotation in fermionic coordinates
Eq.(\ref{ftrans}) would transform an open string field $\Psi$ as
\bea \Big|\Psi, ab \Big\rangle \rightarrow
(\gamma_{g(\alpha)})_{aa'} \Big| g(\alpha) \cdot \Psi, a'b'
\Big\rangle (\gamma^{-1}_{g(\alpha)})_{b'b} \, .
\label{comptransf} \eea
Here, $\alpha = (\alpha_1, \cdots, \alpha_4)$ and $g(\alpha) \cdot
\Psi$ refers to the open string field transformed as Eq.(\ref{ftrans}). Note
that this is consistent with the fact that, as line bundles over
${\mathbb{CP}^3}$, fermionic coordinates are sections of type
$\psi_A \sim {\cal O}(+1)$ and component fields are sections of
type $A \sim 0, \chi \sim {\cal O}(-1), \phi \sim {\cal O}(-2),
\widetilde{\chi} \sim {\cal O}(-3)$ and $G \sim {\cal O}(-4)$.

In the component expansion Eq.(\ref{A}), $A, B, \cdots$ are
indices along fermionic directions in $\mathbb{CP}^{3|4}$ and
hence transform as ${\bf 4}$ under the transformation group SU(4).
This SU(4) is isomorphic to the R-symmetry group in the gauge theory side. It is therefore natural to identify the orbifold action Eq.(\ref{Gamma}) with the
corresponding action in the twistor space:
\bea \psi_A \rightarrow - \psi_A \quad \mbox{for} \quad A = 1,2;
\qquad \psi_B \rightarrow + \psi_B \quad \mbox{for} \quad  B =
3,4. \label{twistororbi} \eea
This is precisely transformations of the sort Eq.(\ref{ftrans}).
Once we identify the ${\mathbb{Z}}_2$ action as in
Eq.(\ref{twistororbi}), it then determines the ${\mathbb{Z}}_2$
action on the components fields according to
Eq.(\ref{comptransf}). Recall that component fields are in
different representations of SU(4), viz.
$\phi_{\overline{i}}^{[AB]}$ transforming in ${\bf 6}$,
$\widetilde{\chi}_{\overline{i}A}$ transforming in $\overline{\bf
4}$, and $G_{\overline{i}}$ transforming as a singlet. After
taking $\gamma_{g(\alpha)}$
 the same as $\gamma_\Gamma$ in Eq.(\ref{gamma}) and quotienting
by the $\mathbb{Z}_2$ action according to the representation
contents under SU(4), we see that the transformation rule
Eq.(\ref{comptransf}) coincides exactly with the orbifold
conditions Eqs.(\ref{proj1}, \ref{proj2}, \ref{proj3}) in the
Yang-Mills theory side. Therefore, component fields surviving the
fermionic orbifold projection in the holomorphic Chern-Simons
theory are precisely the component fields of the quiver gauge
theory surviving the $\Gamma = \mathbb{Z}_2$ orbifold projection
in ${\cal N}=4$ super Yang-Mills theory, as recapitulated in the
previous section.

One can consider orbifold actions by other subgroups of SU(4). The
$\Gamma = \mathbb{Z}_2$ construction worked out above demonstrates
it obvious that a chosen orbifold action acts the same way for the
R-symmetry representations in ${\cal N}=4$ super Yang-Mills theory
and for the fermionic coordinates in the twistor superspace
$\mathbb{CP}^{3|4}$. Equivalently, the orbifold action on the
component fields of D3-brane worldvolume theory is the same as that on
the component fields of ${\cal A}$ in the holomorphic Chern-Simons
theory Eq.(\ref{CS}).

\subsubsection{Linear sigma model on $\mathbb{Z}_2$ super-orbifold}
In the previous subsection, we prescribed topological B-model
on super-orbifold $\mathbb{CP}^{3|4}/\mathbb{Z}_2$. Utilizing
mirror symmetry, one can map it to topological A-model on a mirror
Calabi-Yau superspace.
Identification of the latter would be of interest for
several reasons. First, for the parent theory, it was argued that
topological B-model on $\mathbb{CP}^{3|4}$ is mirror to
topological A-model on a quadric in $\mathbb{CP}^{3|3} \times
\mathbb{CP}^{3|3}$. A feature of this topological A-model as opposed to the
B-model is that parity symmetry \footnote{For the consideration of parity symmetry in topological B-model, see \cite{wittenparity} and references therein.} (under which helicities are
reversed) is manifest \cite{aganagicvafa}.

An interesting question is whether mirror of the topological strings on super-orbifold continues to have a geometric realization as its parent theory does and, if so, whether various discrete symmetries such as the parity are manifest. Second, though the super-orbifold
projection on topological open string sector is well defined, it
is not yet clear to us how to identify topological closed string sector on a
super-orbifold. The reason is because of potential existence of
twisted sectors. A viable route of identifying the twisted sectors
is mapping the topological string to its mirror dual. If the mirror
theory happens to admit a geometric realization on a regular
(super)manifold, then identification of the closed string sector
would be free from such ambiguity. The full spectrum on the mirror can then
be mapped back to that on the original super-orbifold.

With such motivations, we now examine mirror of the super-orbifold
$\mathbb{CP}^{3|4}/\mathbb{Z}_2$. We shall adopt the strategy of
\cite{aganagicvafa}, and describe the super-orbifold in terms of $\sigma$-model. For the covering theory, $\mathbb{CP}^{3|4}$
is described by the U(1) gauged linear $\sigma$-model involving 4
bosonic and 4 fermionic chiral superfields representing the coordinates
 $Z^I$ and $\Theta^I$ of $\mathbb{CP}^{3|4}$, where $I=0,1,2, 3$.
 These fields are subject to the D-term constraint:
\bea \sum_{I=0}^3 |Z^I|^2 + \sum_{I=0}^3 |\Theta^I|^2 = r, \nonumber \eea
where $r$ is the real part of the Fayet-Iliopoulos parameter $r +
i \theta$ and U(1) charges are assigned as
$Q(\Phi^I) = Q(\Theta^I) = 1$ so that the Calabi-Yau condition Str$Q=0$ is obeyed. Adopting the method of \cite{aganagicvafa},
mirror of the topological B-model on $\mathbb{CP}^{3|4}$ can be
constructed by studying the Landau-Ginzburg model whose partition
function is defined by
\bea Z[r] &\equiv& \langle 1 \rangle \nonumber \\
&=& \int \prod_{I=0}^3 \d Y_I \d Z_I \d \eta_I \d \chi_I \, \exp
\Big[ e^{-Y_I} + e^{-Z_I} (1 + \eta_I \chi_I) \Big] \, \delta
\Big(\sum_{J=0}^3 (Y_J - Z_J) - r \Big) \, . \label{motherpart} \eea
Here, the variables in the gauged $\sigma$-model are changed to $Y_I
= |\Phi_I|^2 + \cdots$, $Z_I = - |\Theta_I|^2 +\cdots$, and
$\eta_I, \chi_I$ are a pair of fermionic variables conjugate to
each $Z_I$'s introduced in the course of the change of variables.

The work \cite{aganagicvafa} argued that parity symmetry
of the ${\cal N}=4$ Super Yang-Mills theory is encoded by the
$\mathbb{Z}_2$ symmetry
\bea
{\tt P} \, : \qquad \qquad Y_I & \leftrightarrow & Z_I  \nonumber \\
\eta_I &\rightarrow & e^{-Y_I}  \chi_I  \nonumber  \\
\chi_I &\rightarrow & e^{+Z_I}  \eta_I  \nonumber \\
r &\rightarrow & -r \, , \nonumber \eea
and demonstrated that mirror on which topological A
model is defined by a quadric in $\mathbb{CP}^{3|3} \times
\mathbb{CP}^{3|3}$. On the mirror side, the parity symmetry is
realized as exchange symmetry of the two $\mathbb{CP}^{3|3}$'s and
reversing sign of the moduli $r \leftrightarrow -r$.

Gauged linear $\sigma$-model describing the topological B-model on the
super-orbifold $\mathbb{CP}^{3|4}/\mathbb{Z}_2$ is readily constructed. To this end,
consider a U(1)$\times \widetilde{U(1)}$ gauged linear
$\sigma$-model involving 4 bosonic chiral superfields $\Phi^0,
\cdots, \Phi^3$ and 5 fermionic chiral superfields $\Theta^0,
\cdots, \Theta^4$. We assign the U$(1)\times\widetilde{\rm U(1)}$
charges as $(+1, 0)$ for $\Phi^0, \cdots, \Theta^0, \Theta^1$, as
$(+1,+1)$ for $\Theta^2, \Theta^3$, and as $(0, -2)$ for
$\Theta^4$. The charge assignment is consistent with the
Calabi-Yau condition Str$Q = 0$ and Str$\widetilde{Q}=0$. The
D-term conditions now read
\bea && |\Phi^0|^2 + |\Phi^1|^2 + |\Phi^2|^2 + |\Phi^3|^2 +
|\Theta^0|^2 + |\Theta^1|^2 + |\Theta^2|^2 + |\Theta^3|^2 = r
\label{D1}
\nonumber \\
&& |\Theta^2|^2 + |\Theta^3|^2 - 2 |\Theta^4|^2 = t, \label{D2}
\eea
where $r, t$ are real parts of the two Fayet-Iliopoulos parameters
for $\rm{U(1)} \times \widetilde {\rm U(1)}$. In the limit $t \ll
0$, Eq.(\ref{D2}) indicates that $\vert \Theta^4 \vert^2 $ takes a
large vacuum expectation value, and breaks the $\widetilde{\rm
U(1)}$ gauge group to ${\mathbb{Z}}_2$. Since $\Theta^2, \Theta^3$
are minimally charged under the $\widetilde{\rm U(1)}$ while all
others are neutral, it is seen that $t \ll 0$ limit yields
$\mathbb{CP}^{3|4}/\mathbb{Z}_2$ orbifold, where the
$\mathbb{Z}_2$ acts only on two fermionic coordinates $\Theta^2, \Theta^3$ and none on others.

The corresponding mirror can be constructed by considering the
following one-point correlator within the covering Landau-Ginzburg
theory defined by Eq.(\ref{motherpart}):
\bea Z[r,t] &\equiv& \Big< \int \d Z_4 \d \eta_4 \d \chi_4
\exp\Big[ e^{-Z_4}(1 + \eta_4 \chi_4)\Big] \delta(2Z_4 - Z_2 - Z_3
- t) \Big> \nonumber \\
&=&
\int \prod_{I=0}^3 \d Y_I \prod_{J=0}^4 \d Z_J \d \eta_J \d
\chi_J \exp \Big[ \sum_{I=0}^3 e^{-Y_I} + \sum_{J=0}^4
e^{-Z_J} (1 + \eta_J \chi_J) \Big] \nonumber \\
&\times& \delta \Big(2Z_4 - Z_2 - Z_3 - t \Big) \, \delta
\Big(\sum_{J=0}^3 (Y_J - Z_J) - r \Big). \label{LG2} \eea
The two Fayet-Iliopoulos parameters describe deformation of the
supertwistor orbifold, so $r$ and $t$ ought to originate from
untwisted and twisted sectors, respectively.

Integrating out both $(\eta_0, \chi_0)$ and $(\eta_4, \chi_4)$ and
treating $e^{-Y_0}$ as the Lagrange multiplier $\Lambda$, the
correlator Eq.(\ref{LG2}) is given by
\bea Z[r, t] = \int \d \Lambda \, \Lambda \prod_{i=1}^3 (\d u_i \d
\chi_i  \d v_i \d \eta_i) \exp F(u, v, \eta, \chi) \label{exp1}
\eea
where
\bea F(u,v,\eta,\chi) &=& u_1 v_1 + u_2^2 v_2 + u_3^2 v_3 + u_1 +
u_2^2 + u_3^2 + 1 + \eta_1 \chi_1 + \eta_2 \chi_2 + \eta_3 \chi_3
\nonumber \\
&+& e^r v_1v_2v_3 + e^{-t/2} u_2 u_3. \nonumber \eea
Be it as complicated, the result indicates no apparent geometric
realization of the mirror. Moreover, the correlator is not even
invariant under the parity symmetry, since upon treating
$e^{-X_0}$ as the Lagrangian multiplier $\tilde{\Lambda}$ instead,
the correlator is expressed as
\bea Z[r, t] = \int \d \tilde{\Lambda} \tilde{\Lambda}
\prod_{i=1}^3 (\d u_i \d \chi_i  \d v_i \d \eta_i) u_2^2 u_3^2
v_2^2 v_3^2 \, \exp G(u,v, \eta, \chi) \label{exp2} \eea
where
\bea G(u,v, \eta, \chi) &=& u_1 v_1 + u_2^2 v_2^2  + u_3^2 v_3^2 +
v_1 + v_2^2 + v_3^2 + 1 + \eta_1 \chi_1 + \eta_2 \chi_2 + \eta_3
\chi_3 \nonumber \\
&+& e^{-r} u_1 u_2 u_3 + e^{-t/2} u_2 u_3 v_2 v_3. \nonumber \eea
Comparing the two alternative descriptions Eqs.(\ref{exp1},
\ref{exp2}), we see no sign of a discrete symmetry identifiable
with the parity operation in the gauge theory side.

Though the mirror does not admit geometric realization, one can
still draw a lot of information from the Landau-Ginzburg
description. In case of conventional string theories defined on
ordinary Calabi-Yau spaces, by studying chiral ring structure in
the corresponding Landau-Ginzburg model, cohomological data of the
original Calabi-Yau space could be understood easily. As such, it
would be very interesting to study chiral rings in the above
Landau-Ginzburg model and extract analogous cohomological data of
the super-orbifold $\mathbb{CP}^{3|4}/\mathbb{Z}_2$. Among others,
this may shed light on the field contents arising from the
potential twisted closed string sectors, which by itself is an
important issue in a complete definition of (topological) string
theory on super-orbifolds.
\subsection{${\cal N}=1$ super-orbifold}

Extension of the orbifold construction to theories with lower
supersymmetries is straightforward. Here, we would like to
illustrate the construction for the simplest example yielding ${\cal
N}=1$ supersymmetric quiver gauge theories: supertwistor orbifold
corresponding to the D3-branes localized at the fixed point of
${\mathbb{C}}^3/{\mathbb{Z}}_3$ orbifold.

\subsubsection{Quiver Gauge Theory}
As the orbifold action that would result in ${\cal N}=1$ quiver
gauge theory, we shall consider $\mathbb{Z}_3$ projection on ${\cal
N}=4$ super Yang-Mills theory with gauge group U($3N$), defined by
\bea \Gamma = \left( \begin{array}{ccccccc} \omega &&&&&& \\ && \,\,
\omega &&&& \\ &&&& \,\, \omega && \\ &&&&&& \,\, 1
\end{array} \right) \in \mbox{SU(4)}\eea
and
\bea \gamma_\Gamma = \left( \begin{array}{ccc} {\mathbb{I}}_{N_1} && \\
& \omega {\mathbb{I}}_{N_2} & \\ && \omega^2 {\mathbb{I}}_{N_3}
\end{array} \right) \in \mbox{U}(3N), \nonumber \eea
where $\omega = \exp(2 \pi i /3)$ and $N_1 + N_2 + N_3 = 3N$. The
large-$N$ superconformal invariance condition requires
Tr$\gamma_\Gamma = 0$. This condition is met only for $N_1 = N_2 =
N_3 = N$, so we shall limit ourselves to such a choice of the quiver
gauge group.

Field content of the resulting quiver gauge theory is
determined by orbifold conditions, on which we now work out the
details.
Begin with the gauge bosons. Being SU(4) R-symmetry singlets, they
are subject to orbifold conditions:
\bea \lambda = + \gamma_\Gamma \lambda \gamma^{-1}_\Gamma,
\label{cond1} \eea
where $\lambda$ represents gauge bosons collectively. Explicitly,
express a generic $\lambda$ in color space as
\bea \lambda = \left( \begin{array}{ccc} \lambda_{11} &
\lambda_{12} & \lambda_{13} \\
\lambda_{21} & \lambda_{22} & \lambda_{23} \\
\lambda_{31} & \lambda_{32} & \lambda_{33} \end{array} \right),
\label{matrix}\eea
where each $\lambda_{ij}$ $(i,j=1,2,3)$ constitutes $(N\times N)$
matrix. Then, using
\bea \gamma_\Gamma \lambda \gamma_\Gamma^{-1} = \left(
\begin{array}{ccc} \lambda_{11} & \omega^2 \lambda_{12} &
\omega \lambda_{13} \\
\omega \lambda_{21} & \lambda_{22} & \omega^2 \lambda_{23} \\
\omega^2 \lambda_{31} & \omega \lambda_{32} & \lambda_{33}
\end{array} \right), \nonumber \eea
we solve the orbifold condition Eq.(\ref{cond1}) and find that
only the diagonal entries $\lambda_{11}, \lambda_{22},
\lambda_{33}$ survive as physical degrees of freedom. Therefore,
three sets of $N^2$ gauge bosons survive and the resulting quiver
gauge theory is a theory of gauge group $G=\U(N)_1 \times \U(N)_2
\times \U(N)_3$. Next, consider gauginos $\lambda^1, \cdots,
\lambda^4$. The orbifold conditions are now given by
\bea \lambda^i = \omega \, \gamma \lambda^i \gamma^{-1} \quad
\mbox{for} \quad A=1,2,3; \qquad \lambda^4 = \gamma \lambda^4
\gamma^{-1}. \nonumber \eea
In the notation of Eq.(\ref{matrix}), the surviving components are
$\lambda^A_{12}, \lambda^A_{23}, \lambda^A_{31}$ for $A=1,2,3$ and
$\lambda^4_{11}, \lambda^4_{22}, \lambda^4_{33}$. Finally, for
scalar fields $\lambda^{[AB]}$, the orbifold conditions are given by
\bea \lambda^{[12]} &=& \omega^2 \gamma \lambda^{[12]} \gamma^{-1}
\hskip2cm \lambda^{[13]} = \omega^2 \gamma \lambda^{[13}]
\gamma^{-1} \hskip2cm \lambda^{[23]} = \omega^2 \gamma
\lambda^{[23]} \gamma^{-1} \nonumber \\ \lambda^{[14]} &=& \,
\omega \, \gamma \lambda^{[14]} \gamma^{-1} \hskip2cm
\lambda^{[24]} = \, \omega \, \gamma \lambda^{[24]} \gamma^{-1}
\hskip2cm \lambda^{[34]} = \, \omega \, \gamma \lambda^{[34]}
\gamma^{-1}. \label{c3action} \eea

Putting the surviving components together, We find that field
contents consist of one ${\cal N}=1$ vector multiplet and three
${\cal N}=1$ chiral multiplets transforming as $({\bf N},
\overline{\bf N}, 1), (1, {\bf N}, \overline{\bf N})$ and
$(\overline{\bf N}, 1, {\bf N})$ under the gauge group $\U(N)_1
\times \U(N)_2 \times \U(N)_3$. This is precisely the field
contents of worldvolume theory of D3-branes  on
$\mathbb{C}^3/\mathbb{Z}_3$. As before, Eq.(\ref{c3action})
encodes the orbifold action on $\mathbb{C}^3$, yielding
$\mathbb{C}^3/\mathbb{Z}_3$.
 Denoting the $Z^1, Z^2, Z^3$
the complex coordinates of $\mathbb{C}^3$, the $\mathbb{Z}_3$
action is given by $Z^m \rightarrow \omega Z^m$ ($m=1,2,3$).
Indeed, $\lambda^{[14]}, \lambda^{[24]}, \lambda^{[34]}$
correspond to complex scalars associated with the $Z^1, Z^2, Z^3$
directions, while $\lambda^{[12]}, \lambda^{[13]}, \lambda^{[23]}$
are associated with the complex conjugates (cf. $\phi^{[ij]} = {1
\over 2} \varepsilon^{ijkl} \phi_{[kl]} = (\phi^{[ij]})^*$).

\subsubsection{Orbifolding Topological B-Model}
We would like to understand how the ${\cal N}=1$ quiver gauge
theory can be mapped to an appropriate topological string theory.
Again, we argue that a suitable super-orbifold can be defined
which reproduces correctly the above spectrum of the quiver gauge
theory.

Consider now the holomorphic Chern-Simons theory Eq.(\ref{CS})
with gauge group U(3$N$), so the Chan-Paton indices $a,b$ run over
$1, \cdots, 3N$. Following steps closely parallel to the ${\cal
N}=2$ quiver gauge theory case, we find that an obvious candidate
of the orbifold action pm $\mathbb{CP}^{3|4}$ is
\bea \psi_A \rightarrow \omega \psi_A \qquad \mbox{for} \quad
A=1,2,3; \qquad \psi_4 \rightarrow \psi_4 . \nonumber \eea
This assigns $\mathbb{Z}_3$ action on $\overline{\bf 4}$ of SU(4),
and accordingly determines orbifold action on all other component fields
transforming in different representations of SU(4). One can see
that such orbifold action on the $(0,1)$-form field ${\cal A}$ in
Eq.(\ref{A}) projects the component fields precisely to the same
as the ${\cal N}=1$ vector multiplet of quiver gauge theory with
gauge group $G= \U(N)_1 \times \U(N)_2 \times \U(N)_3$ and three
${\cal N}=1$ chiral multiplets transforming as $({\bf N},
\overline{\bf N}, 1), (1, {\bf N}, \overline{\bf N})$ and
$(\overline{\bf N}, 1, {\bf N})$.

\subsubsection{Linear sigma model on $\mathbb{Z}_3$ super-orbifold}
It is also possible to construct a gauged linear $\sigma$-model
description of the $\mathbb{CP}^{3|4}/\mathbb{Z}_3$
super-orbifold. The sigma model involves four bosonic superfields
$\Phi^0, \cdots, \Phi^3$ and five fermionic superfields $\Theta^0,
\cdots, \Theta^4$. This is the same field contents as the
$\mathbb{Z}_2$ orbifold considered in the previous section. In
order to describe $\mathbb{Z}_3$ super-orbifold, we shall need to
assign $\U(1) \times \widetilde{\U(1)}$ charges $(Q,
\widetilde{Q})$ differently so that the orbifolding now acts on three
fermionic coordinates (instead of two as in the $\mathbb{Z}_2$
case). We assign them as $(+1, 0)$ for $\Phi^0, \cdots, \Phi^3$
and $\Theta^0$, as $(+1,+1)$ for $\Theta^1, \Theta^2, \Theta^3$,
and as $(0, -3)$ for $\Theta^4$. Notice that the Calabi-Yau
conditions, Str$Q = 0$ and Str$\widetilde{Q} = 0$, are satisfied
by such assignment.

The D-term constraints are now given by
\bea && \vert \Phi^0\vert^2 + \vert \Phi^1 \vert^2 + \vert
\Phi^3\vert^2 + \vert \Theta^0\vert^2 + \vert\Theta^1\vert^2 + \vert\Theta^2\vert^2
+\vert\Theta^3\vert^2 = r \nonumber \\
&& \vert \Theta^1 \vert^2 + \vert \Theta^2\vert^2 + \vert
\Theta^3\vert^2 - 3 \vert \Theta_4\vert^2 = t. \eea
The moduli $t$ arises from potential twisted closed string sector.
At $t \rightarrow - \infty$, $\Theta^4$ gets a nonzero vacuum
expectation value, which breaks the $\widetilde{\U(1)}$ gauge
group to ${\mathbb{Z}}_3$. Therefore, the vacuum moduli space is
now reduced to $\mathbb{C}^{3|4}/{\mathbb{Z}}_3$ super-orbifold,
where $\widetilde{\mathbb{Z}}_3$ acts on the fermionic coordinates
as $\Theta^A \rightarrow \omega \Theta^A$ for $A=1,2,3$. Again,
this is just a fermionic counterpart of the
${\mathbb{C}}^3/{\mathbb{Z}}_3$ orbifold and of its linear
$\sigma$-model description.

Partition function of the Landau-Ginzburg $\mathbb{Z}_3$
orbifold is defined as one-point correlator of the parent
Landau-Ginzburg theory:
\bea Z[r, t] &\equiv& \Big< \int \d Z_4 \d \eta_4 \d \chi_4 \,
\exp \Big[ e^{-Z_4} ( 1 + \eta_4 \chi_4) \Big] \delta \Big( 3 Z_4
- Z_1 - Z_2 - Z_3 - t) \Big> \nonumber \\
&=& \int \prod_{I=0}^3 \d Y_I \prod_{J=0}^4 \d Z_J \d \eta_J \d
\chi_J \, \exp \Big[ \sum_{I=0}^3 e^{-Y_I} + \sum_{J=0}^4 e^{-Z_J}
(1 + \eta_J \chi_J) \Big] \nonumber \\
&\times& \delta \Big(3Z_4 - Z_1 - Z_2 - Z_3 - t \Big) \delta \Big(
\sum_{I=0}^3 \Big(Y_I - Z_I) - r \Big). \eea
As in ${\cal N}=2$ super-orbifold case, the one-point correlator
can be expressed in terms of two alternative choices of the
Lagrange multiplier. Comparing the two, we again find no discrete
symmetry identifiable with the parity symmetry in the gauge theory
side.

\subsection{Impossibility of ${\cal N}=3$ super-orbifold}

One might wonder if a variation of the above constructions lead to
a ${\cal N}=3$ super-orbifold. If so, there ought to be some
${\cal N}=3$ gauge theory, dual to topological
string on such super-orbifold. On the other hand, it is known that
the former does not not exist and it actually is equivalent to
${\cal N}=4$ theory. Thus, turned around this way, the
nonexistence of ${\cal N}=3$ (quiver) gauge theory proper may
serve as a check-point for the consistency of the orbifold method
we proposed in this work.

One way of understanding nonexistence of ${\cal N}=3$ quiver gauge
theories is to examine the number of adjoint fermions surviving a
given SU(4) orbifolding. The number of the adjoint fermions are the
fermions satisfying the orbifold condition $\lambda=\gamma \lambda
\gamma^{-1}$. In order to have three such fermions and hence ${\cal
N}=3$ supersymmetry, we should have at least three
singlets for SU(4) elements associated with the orbifold action. The
only such element is the identity of SU(4) and we are forced to go
back to ${\cal N}=4$. In other words, requirement of ${\cal N}=3$
supersymmetry is equivalent to ${\cal N}=4$ supersymmetry.
\subsection{Remarks on non-supersymmetric super-orbifold}
One can also construct non-supersymmetric quiver gauge theories and corresponding topological string theories on appropriate super-orbifolds.
The simplest procedure fitting to the pattern we constructed in the previous sections is to project the ${\cal N}=4$ gauge theory of gauge group U$(4N)$
by the following $\mathbb{Z}_4$ orbifold action defined by simultaneous action on R-symmetry group
\bea \Gamma = \left( \begin{array}{cccc} \omega &&& \\ & \,\,\omega && \\ & & \,\,\omega & \\
&&& \,\,\omega \end{array} \right) \in \mbox{SU(4)} \label{non1}
\eea
and gauge group
\bea \gamma_\Gamma = \left( \begin{array}{cccc} {\mathbb{I}}_{N_1} &&& \\
& \omega {\mathbb{I}}_{N_2} && \\ &&  \omega^2 {\mathbb{I}}_{N_3} & \\
&&&  \omega^3 {\mathbb{I}}_{N_4}
\end{array} \right) \in \mbox{U}(4N). \label{non2}\eea
Here, $\omega = e^{ i \pi \over 2}$ and $N_1 + N_2 + N_3 +N_4=
4N$. Again, to meet the large-$N$ conformal invariance condition, we take
$N_1=N_2=N_3=N_4=N$.

Spectrum of the resulting quiver gauge theory is identified as follows. From the orbifold conditions for gauge bosons:
\bea \lambda=\gamma_\Gamma \lambda \gamma_{\Gamma}^{-1}, \nonumber
\eea
we obtain 4 sets of $N^2$ gauge bosons associated with the gauge group $G = {\rm U}(N)^4$. The adjoint fermions are subject to orbifold conditions:
\bea \lambda^A=\omega\gamma_\Gamma \lambda^A \gamma_{\Gamma}^{-1}
\qquad \mbox{for} \qquad A = 1, \cdots, 4, \nonumber \eea
so we have 4 Weyl fermions transforming bilinearly as $({\bf N},
\overline{{\bf N}},1,1)$, $(1,{\bf N}, \overline{{\bf N}},1)$,
$(1,1,{\bf N}, \overline{{\bf N}})$, $( \overline{{\bf
N}},1,1,{\bf N})$ under four ${\rm U}(N)$ gauge groups. Finally, the
adjoint scalar fields are subject to orbifold conditions:
\begin{equation}
\lambda^{[AB]}=-\gamma_\Gamma \lambda^{[AB]} \gamma_{\Gamma}^{-1},
\end{equation}
leading to six scalars transforming as
$({\bf N}, 1,\overline{{\bf N}},1)$, $(\overline{{\bf N}},1, {\bf
N},1)$, $(1, {\bf N}, 1,\overline{{\bf N}})$ and $(
1,\overline{{\bf N}},1, {\bf N})$. Since all the fermions
transform in `nearest-neighbor' bifundamental representations, they cannot
be paired up to supermultiplets with gauge bosons nor with scalars, and it is evident that the theory is non-supersymmetric.

Repeating the analysis as in previous setions, we find that the relevant
topological string theory is the B-model defined on the super-orbifold $\mathbb{CP}^{3|4}/\mathbb{Z}_4$ obtainable
from the supertwistor space $\mathbb{CP}^{(3|4)}$ by the orbifold action
\bea \psi_A \rightarrow e^{-\frac{i\pi}{2}} \psi_A \qquad
\mbox{for} \qquad A=1, \cdots, 4. \nonumber \eea
Again, it is possible to construct a linear gauged $\sigma$-model
description of the above super-orbifold. The relevant model is the
same one as $\mathbb{Z}_2$ and $\mathbb{Z}_3$ orbifolds, so it
contains four bosonic superfields $\Phi^0, \cdots, \Phi^3$ and
five fermionic superfields $\Theta^0, \cdots, \Theta^4$. To obtain
$\mathbb{Z}_4$ super-orbifold, we would need to assign $\U(1)
\times \widetilde{\U(1) }$ charges so that it now acts on all of the four
fermionic directions. An obvious assignment is as $(+1, 0)$ for
$\Phi^0, \cdots, \Phi^3$, as $(+1,+1)$ for $\Theta^0, \cdots
\Theta^3$, and as $(0, -4)$ for $\Theta^4$, respectively. Notice
that Str$Q$ and Str$\widetilde{Q}$ all vanish, so the resulting
super-orbifold is still a Calabi-Yau manifold. For bosonic
$\mathbb{C}^4/\mathbb{Z}_4$ orbifold, it is known to be a
well-defined Calabi-Yau fourfold, where the orbifold singularity
can be deformed smoothly while preserving the Calabi-Yau
conditions. This is quite analogous in situation to the
$\mathbb{Z}_2$ and $\mathbb{Z}_3$ orbifolds, so we anticipate
that topological B-model on $\mathbb{C}^{3|4}/\mathbb{Z}_4$
super-orbifold would provide twistor description of the above
non-supersymmetric quiver gauge theory.

Another possible choice of the projection leading to a
non-supersymmetric quiver gauge theory is via the action $\omega = -
1$ in Eqs.(\ref{non1}, \ref{non2}), viz. $\mathbb{Z}_2$ orbifold
along all four fermionic directions. The resulting quiver gauge
theory have gauge group U(N)$\times$U(N), containing a pair of
fermions in bifundamental representations and scalars in adjoint
representations. Again, they cannot be organized into
supermultiplets, so the quiver gauge theory is non-supersymmetric.
This model was considered previously in \cite{nekrasov}. Such quiver
gauge theory may opt to define the corresponding super-orbifold,
viz. $\mathbb{CP}^{3|4}/\mathbb{Z}_2$. This model may, however, be
potentially problematic, since we think there would be no gauged
linear $\sigma$-model description obeying Calabi-Yau conditions,
viz. Str$Q$'s all vanish. This follows by inferring from known
results of the bosonic counterpart. For bosonic
$\mathbb{C}^4/\mathbb{Z}_2$ orbifold, it is known that although a
suitable linear sigma-model can be constructed, the Calabi-Yau
conditions, Tr$Q = 0$, are not satisfied because of nonanalytic
behavior of the topological string with respect to K\"ahler
deformations \cite{mohri}. This means that the singular Calabi-Yau
manifold cannot be deformed to a smooth one. It would be interesting
to demonstrate such rigidity directly for the super-orbifold
$\mathbb{CP}^{3|4}/\mathbb{Z}_2$.

\section{Multi-Parton Amplitudes in Quiver Gauge Theories}
Having constructed examples of Witten's gauge-string correspondence for
theories with ${\cal N} < 4$ supersymmetries, in this section, we would like to
test them by comparing multiparton amplitudes computed from both gauge theory and topological B-model. In this section, we shall do so for the simplest
set of them. First, we shall consider MHV multiparton amplitudes involving two fermions. Since the fermions are in bifundamental representations, we expect that such amplitudes would roughly be a product of MHV amplitudes for each product gauge theories. We shall confirm that this expectation is in fact correct.
Second, we shall study MHV multiparton amplitudes involving U(1) gauge
groups. The quiver gauge theory contains overall U(1) subgroup and relative
U(1)'s.  As checkpoints for our orbifold constructions, we will confirm by
explicit computations that overall U(1) decouples while relative U(1) gauge group yields nontrivial multiparton amplitudes exhibiting incoherence.

It is also of phenomenological interest to study multiparton amplitudes
product gauge group $G_1 \times G_2 \times \cdots$ as well as U(1)'s
and fermion or scalar fields in bifundamental representations therein. The Standard Model of the strong and electroweak interactions is certainly of such structure: the gauge group is SU(3)$\times$SU(2)$\times$U(1), and quarks, leptons and Higgs transform in the fundamental or the bifundamental representations. Though the electroweak gauge group is spontaneously broken, at a high-energy regime well above the Fermi scale, it is natural to
expect that multi-parton amplitudes are described by the theory in
the phase where the gauge symmetries are unbroken. At such regime,
all particles can be treated as massless and conformal invariance
would play a role in governing their multi-particle amplitudes. See for example
\cite{yan} for earlier studies.  The ${\cal N}=4$ super Yang-Mills theory contains particles of
helicity $0, \pm 1/2$ and $\pm 1$, all belonging to a single ${\cal
N}=4$ supermultiplet. Therefore, though the corresponding
multi-parton amplitudes involve fermions as well as scalars, these
particles all transform in the adjoint representation of the gauge
group $G$. In order to study particles transforming in other
representations, one needs to relax the supersymmetry from the
maximal ${\cal N}=4$ to lower ones. The quiver gauge theory we
considered in the previous section is close in group structure and
field contents to the Standard Model. In particular, matter fields
transforming in bifundamental representations are readily
obtainable. This suggests that quiver gauge theories may serve as a
laboratory for studying features of multi-particle amplitudes in the
Standard Model.

With such motivations, in this section, we shall study multi-parton
amplitudes of the quiver gauge theories by computing them in the
topological B-model on a chosen super-orbifold. We shall compare
them with known results in gauge theories with product gauge groups, including abelian groups, and with fermions. We shall first recapitulate results regarding such variants, and establish certain extensions relevant for foregoing discussions.

\subsection{Parton Amplitudes for Helicity 0 and 1/2 Adjoint Representations}
The MHV parton amplitude for ${\cal N}=4$ super Yang-Mills theory
with gauge group U($N$) is given by
\cite{parketaylor,nair,manganoreview,witten}
\bea \hat{A}_n = i g^{n-2}_{\rm YM} (2 \pi)^4 \delta^{(4)}(P)
\delta^{(8)} (\Theta) \prod_{i=1}^n {1 \over \langle\la_i, \la_{i+1}\rangle}.
\label{n4amp} \eea
Here, the bosonic and fermionic momenta are
\bea p^i_{a\da} = \la_{ia} \wt \la_{i\da}; \qquad \pi_{aA}^i =
\la_{ia} \eta_{iA} \eea
and $\Theta_{bA}=\sum_i \la_{ib}\eta_{iA}$. We can extract parton
amplitudes for each helicity of the ${\cal N}=4$ vector multiplet by
expanding the Dirac delta function for the total fermionic momentum.
For example, we obtain in this way the MHV amplitudes involving two
external fermions and $n-2$ gluons as
\bea \hat{A}_n = i g^{n-2}_{\rm YM} (2 \pi)^4 \delta^{(4)}(P)
\langle q\alpha\rangle^3\langle q'\alpha\rangle
\prod_{i=1}^n {1 \over \langle\la_i, \la_{i+1}\rangle}.
\eea
Here, the fermion denoted by $q$ has the helicity $-\frac{1}{2}$ and
the fermion $q'$ has the helicity $\frac{1}{2}$ while the gluon
denoted by $\alpha$ has the helicity $-1$. Similarly, we obtain MHV
amplitudes involving 4 scalars of helicity 0 as
\bea \hat{A}_n = i g^{n-2}_{\rm YM} (2 \pi)^4 \delta^{(4)}(P)
(\langle 13\rangle\langle 32\rangle\langle 24\rangle\langle 41\rangle
+(2 \leftrightarrow 4)) \prod_{i=1}^n {1 \over
\langle\la_i, \la_{i+1}\rangle}, \eea
where we denoted four scalars by $i=1, \cdots, 4$.

The MHV amplitudes in component form can be obtained in twistor
space as well. Fourier transforming Eq.(\ref{n4amp}), the MHV parton
amplitude in twistor space is given by
\bea \wt{A}_n (\la^a_i, \mu^\da_i, \psi^A_i) = i g^{n-2}_{\rm YM}
\int \d^4 x_{\a\da} \d^8 \th_\a^A \, \prod_{i=1}^n \delta^{(2)}
(\mu_{i \da} + x_{\a\da} \la_i^\a) \delta^{(4)} (\psi^A_i +
\th_\a^A \la_i^\a) \prod_{j=1}^n {1 \over \langle \la_j, \la_{j+1}
\rangle }. \eea
By expanding the fermionic coordinates and picking up suitable
terms, one can obtain the MHV amplitudes involving various component
fields.

\subsection{Multi-Parton Amplitudes for Quiver Gauge Groups}
Consider a quiver-type gauge theory with product gauge group
SU($N_1$)$\times$SU($N_2$). Consider also quarks which transform
in the bi-fundamental representation $({\bf N}_1, \overline{{\bf
N}_2})$ under these gauge groups and their complex conjugates. Then,
the full multi-parton amplitudes ${\cal M}$ involving a
quark-antiquark pair with $n_1$ gauge bosons of SU($N_1$) and
$\widetilde{n}_2$ gauge bosons of SU($N_2$) can be written as
\cite{manganoreview}
\bea {\cal M}(q, 1, \cdots, n_1; \wt{1}, \cdots, \wt{n}_2,
\overline{q}) &=& \sum_{P(n_1), P(n_2)} (X^1 \cdots X^{n_1})_{ij}
(Y^1 \cdots
Y^{n_2})_{\overline{j} \overline{i}} \nonumber \\
&\times& {\cal A}_{N_1, N_2} (\overline{q}, 1, \cdots, n_1, q; q,
\wt{1}, \cdots, \wt{n}_2, \overline{q}). \label{prodamp} \eea
Here, the sum is over all permutations of $n_1$ gauge bosons of
SU$(N_1$) gauge group between the quark and the antiquark and
similarly of $\widetilde{n}_2$ gauge bosons of SU($N_2$) gauge group between the antiquark and the quark. The $ij, \overline{ji}$ indices refer to
SU($N_1$) and SU($N_2$) `color' indices of the quark-antiquark
pair. $X^A$, $Y^B$ are generators of SU$(N_1)$ and SU$(N_2)$ gauge
groups, respectively.

The sub-amplitudes ${\cal A}_{N_1, N_2}$ defined as in
Eq.(\ref{prodamp}) are readily obtained in terms of the
multi-parton amplitudes ${\cal A}$ involving a quark-antiquark
pair:
\bea {\cal A}_{N_1, N_2} (\overline{q}, 1, \cdots, n_1, {q}; q,
\wt{1}, \cdots, \wt{n}_2, \overline{q}) = \sum_{I} {\cal A}({q},
1, \cdots, n_1; \wt{1}, \cdots, \wt{n}_2, \overline{q}),
\label{subamprel} \eea
where the sum over $I$ refers to over all possibilities the gauge
bosons of the second gauge group can be interspersed within those
of the first gauge group maintaining the order of both the first
and the second set of gauge bosons. This sum renders all Feynman
diagrams which connect directly the gauge bosons of SU($N_1$) with
those of SU($N_2$) to be cancelled.

Concretely, consider the scattering in which the quark has $-$
helicity and the gauge boson $\alpha$ of either gauge group has
$-$ helicity while all other partons have $+$ helicities. Then,
the corresponding multi-parton amplitude is given by
\bea {\cal A}_{N_1, N_2} (\overline{q}, 1, \cdots, n_1, {q}; q,
\wt{1}, \cdots, \wt{n}_2, \overline{q}) &=& i {\langle q \alpha
\rangle^3 \langle \overline{q} \alpha \rangle \over \langle q
\overline{q} \rangle^2} \sum_I {\langle \overline{q} q \rangle
\over \langle \overline{q} 1 \rangle \langle 1 2 \rangle \cdots
\langle n_1 \wt{1} \rangle \langle
\wt{1} \wt{2} \rangle \cdots \langle \wt{n}_2 \overline{q} \rangle } \nonumber \\
&=& i {\langle q \alpha \rangle^3 \langle \overline{q} \alpha
\rangle \over \langle \overline{q} q \rangle^2} {\langle
\overline{q} q \rangle \over \langle \overline{q} {1} \rangle
\langle {1} {2} \rangle \cdots \langle n_1 q \rangle} {\langle
\overline{q} q \rangle \over \langle q \wt{1} \rangle \langle
\wt{1} \wt{2} \rangle \cdots \langle \wt{n}_2 \overline{q}
\rangle},  \label{field} \eea
where in the second expression we used the relation
Eq.(\ref{subamprel}). In subsection 3.4, we will reproduce this
amplitude Eq.(\ref{field}) directly from the amplitudes which
descends from topological B-model on $\mathbb{CP}^{3|4}$ by
taking the super-orbifold projections we identified in previous
sections.

\subsection{U(1) Gauge Group}
The multi-parton amplitudes involving abelian gauge group are also
of interest from various viewpoints. First, quiver gauge theories
constructed out of D3-branes contain U(1) subgroups. Among these,
overall U(1) subgroup decouples from the rest. The decoupling is obvious
from D3-brane viewpoint, but confirmation of such decoupling within
the multiparton amplitudes would constitute an interesting checkpoint
of the super-orbifold prescription proposed in this work. The
rest are relative U(1) groups, and they have nontrivial
multiparton amplitudes involving charged particles. Derivation of these
amplitudes from topological B-model would offer another interesting checkpoint. Second, from phenomenological viewpoint, the
Standard Model contains U(1) hypercharge interactions, and it is
of interest to model high-energy scattering mediated by U(1)
hypercharge interactions via multiparton amplitudes of the
relative U(1) groups. With such motivations, we consider U(1)
multiparton amplitude involving a pair of quark-antiquark.

The multi-parton amplitudes for U(1) gauge group is extremely
simple, and they are obtainable by setting the gauge group
generators to those for the U(1) subgroup of interest. Thus, the
resulting MHV amplitude is given by
\bea {\cal A}_{\rm U(1)} (q, 1, \cdots, n, \overline{q}) =
\sum_{P} {\cal A} (q, 1, \cdots, n, \overline{q}), \nonumber \eea
where the sum runs over {\sl all} permutations of the gauge
bosons. It has the remarkable effect of causing all nonabelian
Feynman diagrams to cancel one another.

Consider the scattering of $n$ photons with a quark-antiquark pair
in which the quark and the photon labeled as $\alpha$ have $-$
helicity and all other partons have $+$ helicity. The amplitude is
then described by
\bea {\cal A}_{\rm U(1)} (q, 1, \cdots, n, \overline{q}) &=& i
{\langle q \alpha \rangle^3 \langle \overline{q} \alpha \rangle
\over \langle q \overline{q} \rangle^2} \sum_P { \langle
\overline{q} q \rangle \over \langle q 1 \rangle \langle 1 2
\rangle \cdots \langle n \overline{q} \rangle} \nonumber \\
&=& i {\langle q \alpha \rangle^3 \langle \overline{q} \alpha
\rangle \over \langle q \overline{q} \rangle^2} \prod_i {\langle
\overline{q} q \rangle \over \langle q i \rangle \langle i
\overline{q} \rangle}. \label{relativeu1} \eea
Here, in the second expression, we have used the eikonal-like identity:
\bea \sum_{P_n} {\langle p \overline{p} \rangle \over \langle p 1
\rangle \langle 1 2 \rangle \cdots \langle n \overline{p} \rangle
} = \prod_{a=1}^n {\langle p \overline{p} \rangle \over \langle p a
\rangle \langle a \overline{p}\rangle}, \label{identity} \eea
where we emphasize that the permutation $P_n$ involves only $n$
particles labeled by $1, 2, \cdots, n$. This identity follows
straightforwardly from repeated use of the Fierz identity: $\langle
a \overline{a} \rangle \langle b \overline{b} \rangle = \langle a
\overline{b} \rangle \langle b \overline{a}\rangle + \langle a b
\rangle \langle \overline{a} \overline{b} \rangle$. The eikonal-like
form of the amplitude Eq.(\ref{relativeu1}) reflects the physics
that multiple soft photons are emitted incoherently one another.

It is then straightforward to combine this result with that of the
previous subsection. Consider the gauge group
SU$(N_1)\times$SU($N_2)\times$U(1), and this can be viewed as
U$(N_1)\times$U($N_2)/{\rm U}_d(1)$ where the quotient
is the overall U(1) group. Consider MHV amplitudes for a
quark-antiquark pair scattering with $n_1$ SU($N_1$) gauge bosons,
$n_2$ SU$(N_2$) gauge bosons, and $n$ U(1) photons. It is
straightforward to express the amplitude as
\bea && {\cal A}_{N_1, N_2, \rm{U}(1)} (\overline{q}, 1 \cdots,
n_1, {q}; q, \wt{1}, \cdots, \wt{n}_2, \overline{q}; q, \wh{1},
\cdots, \wh{n}, \overline{q}) \nonumber \\
&=& i {\langle q \alpha \rangle^3 \langle \overline{q} \alpha
\rangle \over \langle q \overline{q} \rangle^2}
{\langle \overline{q} q \rangle \over \langle \overline{q} 1
\rangle \langle 1 2 \rangle \cdots \langle n_1 {q} \rangle }
{\langle \overline{q} q \rangle \over \langle q \wt{1} \rangle
\langle \wt{1} \wt{2} \rangle \cdots \langle \wt{n}_2 \overline{q}
\rangle }
\prod_{a=1}^n {\langle \overline{q} q \rangle \over \langle q
\wh{a} \rangle \langle \wh{a} \overline{q} \rangle}. \label{total}
\eea
Here, as above, we assigned the helicities so that the quark $q$
and $\alpha$-th gauge boson (out of $n_1 + n_2 + n$ gauge bosons
involved) carry $-$ helicity while all other particles carry $+$
helicities.

In the following subsection, from the topological string
amplitudes on super-orbifold $\mathbb{CP}^{3|4}/\mathbb{Z}_2$, we
will reproduce these amplitudes involving U(1) subgroups as well.

\subsection{U($N)\times$U($N$) Parton Amplitudes from Topological
B-Model Orbifold}
Having constructed topological strings on super-orbifolds that
correspond to quiver gauge theories with product gauge groups and
lower supersymmetries, parton amplitudes can be computed
straightforwardly from Witten's topological B-model amplitudes, which are
the same as ${\cal N}=4$ super Yang-Mills multiparton amplitudes, by
taking into account of appropriate orbifold projections.

We will now compute some of such amplitudes, in
particular, those involving bi-fundamental fermions and show that
they reproduce those gauge theory amplitudes summarized in the
previous subsections. This is actually a simple matter once we show
that such amplitudes can be obtained from ${\cal N}=4$ multiparton
amplitudes with a suitable orbifold projection. Since the
topological strings on $\mathbb{CP}^{(3|4)}$ reproduces the ${\cal
N}=4$ multiparton amplitudes and since the same orbifold action was
used for defining the corresponding topological strings, it follows
that the multiparton amplitudes involving bi-fundamental fermions
could be obtained from the topological strings on a suitable
super-orbifold.

To demonstrate this,  we take the ${\cal N}=2$ quiver gauge theory
considered in the previous section. After the $\mathbb{Z}_2$
orbifold projection, the gauge groups were U$(N)_1
\times$U$(N)_2$. We will begin with amplitudes involving
nonabelian gauge bosons. Denote in matrix notation the gauge
bosons of SU$(N)_1$ group by
\bea
\lambda_1 \Gamma \qquad \mbox{where} \qquad \Gamma = \left(
\begin{array}{cc}
         X & 0  \\  0 & 0
       \end{array}  \right)   \label{X}
\eea
and gauge bosons of SU$(N)_2$ group by
\bea \lambda_2
\overline{\Gamma} \qquad \mbox{where} \qquad \overline{\Gamma} =
\left(
\begin{array}{cc}
         0 & 0  \\  0 & Y
       \end{array}  \right)    \label{Y}
\eea
The quarks transforming as $(N, \overline{N})$ are denoted by
\bea q Q \qquad \mbox{where} \qquad Q = \left(
\begin{array}{cc}
          0 & W \\  0 & 0
       \end{array}  \right),     \label{Q}
\eea
while antiquarks transforming as $( \overline{N}, N)$ are denoted
by
\bea
\overline{q} \overline{Q} \qquad \mbox{where} \qquad \overline{Q} =
\left(
\begin{array}{cc}
          0 & 0 \\  W^\dagger & 0
       \end{array}  \right).    \label{Qbar}
\eea
Rather than working directly with the topological B-model amplitudes,
since they are the same as multiparton amplitudes in ${\cal N}=4$
supersymmetric gauge theory, we will work with the latter. The multi-parton amplitudes in the quiver gauge theory are then obtainable straightforwardly from the covering ${\cal N}=4$
multi-parton amplitudes in which all of these particles correspond
to various components of the parent gauge boson. Thus, it is a
simple matter to obtain the MHV amplitude with two external
fermions and $m+n$ gluons, where $m, n$ are the number of gauge
bosons of SU$(N)_1$ and SU$(N)_2$ groups, respectively. We find,
stripping off the gauge coupling constant and overall momentum
conserving delta functions,
\bea {\cal A}_{\rm quiver} &\equiv& i \sum_{P_{m+n+2}} \mbox{Tr}
(\Gamma_1\Gamma_2 \cdots \Gamma_m Q
\overline{\Gamma}_{{1}}\overline{\Gamma}_{{2}} \cdots
\overline{\Gamma}_{{m}}\bar{Q}) \nonumber \\
\hskip1cm &\times& \frac{\langle
q\alpha\rangle^3\langle\overline{q}\alpha\rangle}{
\langle\overline{q}1\rangle \langle 12\rangle \langle 23\rangle
\cdots \langle mq\rangle \langle q \wt{1} \rangle \langle \wt{1}
\wt{2} \rangle \cdots \langle \wt{n} \overline{q} \rangle}.
\label{quiveramp}\eea
Here, $P_{m+n+2}$ descends directly from the definition of
multiparton amplitudes in the covering ${\cal N}=4$ super
Yang-Mills theory, and hence involves permutations of all $m+n+2$
partons. However, one readily observes that permutations directly
connecting gauge bosons of SU$(N)_1$ and SU$(N)_2$ vanishes upon
taking the color trace. Thus the resulting expression reduces
precisely to the same one as the gauge theory result Eq.(\ref{prodamp}),
where the kinematical part of the amplitude agrees precisely with
the second expression in Eq.(\ref{field}).

A comment is in order. The MHV amplitudes we derived in
Eq.(\ref{quiveramp}) actually holds for SU$(N_1)\times$SU$(N_2)$
for arbitrary $N_1, N_2$ with $N_1 + N_2 = 2N$. This is because
the conformality condition Tr$\gamma_\Gamma =0$ enters beginning
at one-loop order. Thus, agreement of Eq.(\ref{quiveramp}) with
Eq.(\ref{field}) is exact for all tree-level MHV amplitudes.

We can also examine multiparton amplitudes for U(1) subgroups.
This amounts to replacing the generators for gauge bosons and
quark-antiquark pair by those describing a U(1) subgroup of
interest. Thus, for overall U(1) subgroup, which we know to be decoupled
and would yield vanishing amplitude, we would replace
$\Gamma, \overline{\Gamma}$ in Eqs.(\ref{X}, \ref{Y}) by
\bea \Gamma, \overline{\Gamma} \rightarrow \left( \begin{array}{cc} \mathbb{I} & 0 \\
0 & \mathbb{I} \end{array} \right). \nonumber \eea
Then, the color factor is reduced to an overall constant Tr$W
W^\dagger$ and picks up the color indices of quark-antiquark pair, so the amplitude is reduced to sum over permutations
of kinematical amplitude in Eq.(\ref{quiveramp}). Utilizing the eikonal-like
identity Eq.(\ref{identity}), we can express
\bea \sum_{P_m, P_n} {1 \over \langle \overline{q} 1 \rangle
\cdots \langle m q \rangle \langle q \wt{1} \rangle \cdots \langle
\wt{n} \overline{q} \rangle} = (-)^n \prod_{a=1}^{m+n} {\langle q
\overline{q} \rangle \over \langle q a \rangle \langle a
\overline{q} \rangle}. \label{partial} \eea
Notice the sign factor $(-)^n$ in the final expression. It originates from reversing the order
$\langle q 1 \rangle \cdots \langle n \overline{q} \rangle $ into
$\langle \overline{q} n \rangle \cdots \langle 1 q \rangle$ so
that the identity Eq.(\ref{identity}) can be applied uniformly for
all $m+n$ U(1) gauge bosons. Since $\Gamma = \overline{\Gamma}$,
we will need to consider other amplitudes proportional to the
color factor:
\bea \mbox{Tr} (\Gamma_1 \cdots \Gamma_p Q \overline{\Gamma}_1
\cdots \overline{\Gamma}_q \overline{Q}) \qquad \mbox{for} \quad
(p+q) = (m+n) \eea
which all result in the same form of the amplitude. Summing
Eq.(\ref{partial}) over all such possibilities, the sign factor
$(-)^n$ conspires with the combinatorial factors such that the
total sum vanishes identically. This confirms our anticipation
that the overall U(1) subgroup decouples directly at the level of
multiparton amplitudes. Notice that the sign factor $(-)^n$, which originates
from the kinematical part of the amplitude was crucial for leading to such
a nullifying result.

For the relative U(1) gauge subgroup, the generators for gauge
bosons are now to be replaced by
\bea \Gamma, \overline{\Gamma} \rightarrow \left( \begin{array}{cc} \mathbb{I} & 0 \\
0 & -\mathbb{I} \end{array} \right). \nonumber \eea
Again, the color factor is reduced to an overall constant, but now
its value is weighted by an extra sign factor and yields $(-)^n
\mbox{Tr} W W^\dagger$. More specifically, the extra sign factor
$(-)^n$ arises as one passes $Q$ through $\overline{\Gamma}_1
\cdots \overline{\Gamma}_n$ and place it next to $\overline{Q}$.
Thus, in repeating the same combinatorial considerations, this
sign factor cancels the sign factor $(-)^n$ that originated from
the kinematical amplitudes. As such, upon summing over all
amplitudes involving $(m+n)$ gauge bosons, we obtain a
nonvanishing amplitude, which is the same as gauge theory result
Eq.(\ref{relativeu1}), up to relative normalization of gauge coupling
contants.

It is also straightforward to put all these together, and it is
readily seen that the most general multiparton amplitudes
involving quark-antiquark pair as obtained from topological
strings theory via super-orbifolding agrees with the field theory
result Eq.(\ref{total}).

\section*{Acknowledgement}
We thank J. Maldacena and E. Witten for useful discussions at early stage of this work.

\end{document}